# PERFORMANCE OF ELECTRICAL SPECTROSCOPY USING A RESPER PROBE TO MEASURE SALINITY AND WATER CONTENT OF CONCRETE AND TERRESTRIAL SOIL

A. Settimi[(1)]*

[(1)] Istituto Nazionale di Geofisica e Vulcanologia, Sezione di Geomagnetismo, Aeronomia e Geofisica Ambientale, Via di Vigna Murata 605, I-00143 Rome, Italy

*Short Title:* **RESPER PROBE TO MEASURE SALINITY AND WATER CONTENT**

**\*Corresponding author: Dr. Alessandro Settimi**

Tel: +39-065-1860719

Fax: +39-065-1860397

Email: alessandro.settimi@ingv.it

**Abstract**

This report discusses the performance of electrical spectroscopy using a resistivity/permittivity (RESPER) probe to measure salinity $s$ and volumetric content $\theta_W$ of water in concrete and terrestrial soil. A RESPER probe is an induction device for spectroscopy which performs simultaneous noninvasive measurements of electrical resistivity $1/\sigma$ and relative dielectric permittivity $\varepsilon_r$ of a subjacent medium. Numerical simulations show that a RESPER probe can measure $\sigma$ and $\varepsilon$ with inaccuracies below a predefined limit (10%) up to the high frequency band. Conductivity is related to salinity, and dielectric permittivity to volumetric water content using suitably refined theoretical models that are consistent with predictions of the Archie and Topp empirical laws. The better the agreement, the lower the hygroscopic water content and the higher the $s$; so closer agreement is reached with concrete containing almost no bonded water molecules, provided these are characterized by a high $\sigma$. The novelty here is application of a mathematical–physical model to the propagation of measurement errors, based on a sensitivity functions tool. The inaccuracy of salinity (water content) is the ratio (product) between the conductivity (permittivity) inaccuracy, as specified by the probe, and the sensitivity function of the salinity (water content) relative to the conductivity (permittivity), derived from the constitutive equations of the medium. The main result is the model prediction that the lower the inaccuracy of the measurements of $s$ and $\theta_W$ (decreasing by as much as an order of magnitude, from 10% to 1%), the higher the $\sigma$; so the inaccuracy for soil is lower. The proposed physical explanation is that water molecules are mostly dispersed as $H^+$ and $OH^-$ ions throughout the volume of concrete, but are almost all concentrated as bonded $H_2O$ molecules only at the surface of soil.

# 1. Introductory review

## 1.1. Electrical spectroscopy

Electrical resistivity and relative dielectric permittivity are two independent physical properties that characterize the behavior of bodies when they are excited by an electromagnetic field. The measurement of these properties provides crucial information regarding the practical use of the bodies (for example, materials that conduct electricity), as well as for numerous other purposes.

Some studies have shown that the electrical resistivity and dielectric permittivity of a body can be obtained by measuring the complex impedance using a system with four electrodes, although these electrodes do not require resistive contact with the investigated body (Grard, 1990a, b; Grard and Tabbagh, 1991; Tabbagh et al., 1993; Vannaroni et al., 2004; Del Vento and Vannaroni, 2005). In this case, the current is made to circulate in the body by electric coupling, by supplying the electrodes with an alternating electrical signal of low (LF) or middle (MF) frequency. In this type of investigation, the range of optimal frequencies for electrical resistivity values of the more common materials is between ~10 kHz and ~1 MHz.

The lower limit is effectively imposed by two factors: a) First, the Maxwell-Wagner effect, which limits probe accuracy (Frolich, 1990). This is the most important limitation and occurs because of interface polarization effects that are stronger at low frequencies, e.g., below 10 kHz, depending on the medium conductivity; b) Secondly, the need to maintain the amplitude of the current at measurable levels, because with the capacitive coupling between electrodes and soil the current magnitude is proportional to the frequency.

Conversely, the upper limit is fixed so as to allow analysis of the system under a regime of quasi-static approximation, ignoring the factor of the velocity of the cables used for the electrode harness, which worsens the accuracy of the impedance phase measurements. It is therefore possible to make use of an analysis of the system in the LF and MF bands where the electrostatic term is significant. A general electromagnetic calculation produces lower values than a static one, and high resistivity reduces this difference. Consequently, above 1 MHz, a general electromagnetic calculation must be preferred, while below 500 kHz, a static calculation would be preferred; between 500 kHz and 1 MHz, both of these methods can be applied (Tabbagh et al., 1993).

Unlike a previous study (Tabbagh et al., 1993), the present numerical simulations show that the upper frequency limit can be raised to around 30 MHz. The agreement between the two calculations is excellent at MFs, and only small differences are seen at high frequencies (HFs) for the imaginary part relative to the real part of the complex impedance.

## 1.2. Salinity and volumetric water content

Volumetric water content is a key variable in hydrological modeling. Monitoring water content in the field requires a rapid and sufficiently accurate method for repetitive measurements at the same location (Schön, 1996).

Most of the main disadvantages of radiation techniques do not occur using methods in which volumetric water content is established from the dielectric properties of wet media. Relative dielectric permittivity is generally defined as a complex entity. However, in the present report, dielectric permittivity refers only to the real part. The imaginary part of permittivity stems mainly from electrical conductivity and can be used to assess salinity (Archie, 1942; Corwin and Lesch, 2005a, b). The permittivity of a material is frequency dependent, and so the sensitivity of these methods is also frequency dependent.

Understanding the relationships between the effective permittivity of concrete and terrestrial soil $\varepsilon$ and their water contents $\theta_W$ is important, because measurements of effective permittivity are used to establish moisture content. This report addresses the $\varepsilon(\theta_W)$ relationship in the HF band, from a few MHz to around 30 MHz, which is relevant for determining the moisture content in porous media. An unsaturated porous medium is considered as a three-component mixture of solids, water and air, each of which has significantly different permittivities: 5, 80, and 1, respectively.

While the water content in a concrete or soil mixture is usually much less than the volume of aggregates, it makes the main contribution to the complex permittivity of the overall mixture. This is because the permittivity of water is much higher than that of the other components. Furthermore, the electromagnetic property of water is strongly influenced by the quantity of dissolved salts. Therefore, a portion of the present report is focused on modeling the dielectric properties of saline (Klein and Swift, 1977).

*Concrete* (Schön, 1996). Volumetric water content, salinity and porosity affect the relative dielectric permittivities of porous construction materials, like concrete and masonry (Cheeseman et al., 1998; Gonzalez-Corrochano et al., 2009). These materials are classified as heterogeneous mixtures and they are typically comprised of two or more components that have considerably different dielectric properties. This report discusses a number of dielectric mixing models that were applied to estimate the effective dielectric properties of matured concrete. These models are often known as 'forward models', because they start from a basis of assumed proportions and spatial distributions of components of known dielectric permittivity.

Many types of dielectric models have been developed to cover a wide range of circumstances (not related to concrete), and several comprehensive reviews of the topic are available in the literature (e.g., Robert, 1998). For the purposes of the present study, these can be broadly divided into simple volumetric models and geometric dielectric models (Halabe et al., 1993; Tsui and Matthews, 1997).

*Terrestrial soil* (Schön, 1996). Two different approaches have been taken when relating volumetric water content to relative dielectric permittivity. In the first approach, functional relationships are selected purely for their mathematical flexibility in fitting the experimental data points, with no effort being made to provide physical justification (Banin and Amiel, 1970). Various empirical equations have been proposed for the relationship between $\varepsilon$ and $\theta_W$. The most commonly used equation (Topp et al., 1980) is suggested to be a valid approximation for all types of mineral soils. This and other equations have been shown to be useful for most mineral soils, although they cannot be applied to all types of soil, e.g., peat and heavy clay soils, without calibration.

In the second approach, the functional form of the calibration equation is derived from dielectric mixing models that relate the composite dielectric permittivity of a multiphase mixture to the permittivity values and volume fractions of its components, on the basis of the assumed geometrical arrangement of the components (De Loor, 1964; Sen et al., 1981; Carcione et al., 2003).

To better understand the dependence of permittivity on water content, porosity $\eta$, and other characteristics of porous media, it is necessary to resort to physically based descriptions of two-phase and three-phase mixtures (Roth et al., 1990). To characterize this dependence on large-surface-area materials, it was proposed that another component be included: bonded water, with much lower permittivity than free water (Friedman, 1998; Robinson et al., 2002).

### 1.3. Structure of the report

Following this introductory review, section 2 defines salinity and porosity, giving typical values for both concrete and terrestrial soil. Section 3 discusses the dielectric properties of water and refines the model that describes the relative dielectric permittivity of water as a function of the distance from the soil surface. Section 4 recalls some empirical and theoretical models that link electrical conductivity to porosity, and introduces the function of sensitivity for the conductivity relative to salinity. Section 5 reiterates some empirical and theoretical models that link dielectric permittivity to volumetric water content, and introduces the sensitivity function (Murray-Smith, 1987) of permittivity relative to volumetric water content for both concrete and soil. Section 6 describes the RESPER probe, as connected to an analog-to-digital converter (ADC), which samples in phase and quadrature (IQ) mode (Jankovic and Öhman, 2001), and calculates the established inaccuracies in the measurements of conductivity and permittivity. Section 7 applies the sensitivity function method for calculating inaccuracies in measurements of salinity and water content established by the RESPER probe. Section 8 presents the conclusions. Finally, the Appendix provides an outline of the somewhat lengthy calculations that are required.

## 2. Salinity and porosity

The salinity $s$ of a salt solution is defined as the total solid mass in grams of salt that are dissolved in 1.0 kg of an aqueous solution. Salinity is therefore expressed in parts per thousand (ppt) by weight. The term $s$ represents the total of all of the salts dissolved in the water, in terms of the sodium chloride (NaCl) equivalent (Corwin and Lesch, 2005a, b). The salinity $s$ of pore water in concrete and terrestrial soil is generally much smaller than 10 ppt.

The loose bulk density ($\rho_b$, expressed in g/cm$^3$) is calculated as the $W/V$ ratio, where $W$ is the weight of the aggregates inside a recipient of volume $V$ (Gonzalez-Corrochano et al., 2009; Banin and Amiel, 1970).

The particle density (apparent and dry, expressed in g/cm$^3$) is determined using an established procedure described by Gonzalez-Corrochano et al. (2009). According to this standard:

- The apparent particle density $\rho_a$ is the ratio between the mass of a sample of aggregates when dried in an oven, and the volume that the aggregates occupy in water, including internal water-tight pores and excluding pores open to water.

- The dry particle density $\rho_p$ is the ratio between the mass of a sample of aggregates when dried in an oven, and the volume that the aggregates occupy in water, including internal water-tight pores and pores open to water.

The porosity $\eta$ (air-filled space between aggregates in a container) is calculated using the established method described by Gonzalez-Corrochano: $\eta=1-\rho_b/\rho_p$ where $\eta$ is the void percentage (%), $\rho_b$ is the loose bulk density, and $\rho_p$ is the dry particle density, of the sample.

Cement paste porosity depends fundamentally on the initial water-to-cement (*W/C*) ratio and the degree of cement hydration. The relationship between porosity and cement paste processing was extensively investigated by Cheeseman et al. (1998). Pressed cement paste samples that contained no waste additions and had initial *W/C* ratios of 0.4 and 0.5 were prepared. Pressing at 16 MPa reduced the *W/C* of these samples to less than half their initial values. Increasing the pressure to 32 MPa further reduced the final *W/C* ratios.

Fine textured terrestrial soils that are characterized by a bulk density of $\rho_b = 1.2$ g/cm$^3$, and coarse textured soils, with $\rho_b = 1.6$ g/cm$^3$, have been studied (Friedman, 1998). The particle densities of the soils and of pure clay minerals, $\rho_p$ (required for calculating porosity $\eta$), is assumed to be 2.65 g/cm$^3$, unless another value is known. For the soils from Dirksen and Dasberg (1993), which contained small amounts of organic matter (up to 5%), the particle densities were estimated to be $\rho_p$ (g/cm$^3$) = 2.65 × % minerals + 1.0 × % OM, where OM was the organic matter.

## 3. The dielectric properties of water

While the volumetric fraction of water in a mixture is small, it nevertheless has a very marked effect on the velocity and attenuation of the electromagnetic waves in concrete or terrestrial soil, because of its high complex relative dielectric permittivity. This property of water is strongly influenced by the presence of dissolved salts. Only salts that are actually in solution at any given time will affect the dielectric properties of water, and of the mixture as a whole. The presence of dissolved salts slightly reduces the real part of the complex dielectric permittivity of water (which increases the wave velocity), and greatly increases the imaginary part (which increases the attenuation of electromagnetic waves). This latter effect is due to the increased electrical conductivity of the water. Furthermore, the temperature $t$ of water affects its conductivity, and this is another factor that influences its dielectric properties, which are also a function of the frequency $f$ of the electromagnetic waves (Klein and Swift, 1977).

The complex permittivity of sea water can be calculated at any frequency within the HF band using the Debye (1929) expression, which in its most general form, is given by:

$$\varepsilon_W^{(C)}(t,s,\omega) = \varepsilon_W^{(R)}(t,s,\omega) + i\varepsilon_W^{(I)}(t,s,\omega) =$$
$$= \varepsilon_\infty + \frac{\varepsilon_{stat}(t,s) - \varepsilon_\infty}{1 + [j\omega \cdot \tau(t,s)]^{1-\alpha}} - i\frac{\sigma_{stat}(t,s)}{\omega\varepsilon_0},$$

where $\omega$ is the angular frequency (in rad/s) of the electromagnetic wave (= $2\pi f$, with $f$ as the cyclic frequency in Hz), $\varepsilon_\infty$ is the relative dielectric permittivity at infinite frequency, $\varepsilon_{stat}$ is the static dielectric permittivity, $\tau$ is the relaxation time in s, $\sigma_{stat}$ is ionic or ohmic conductivity, which is sometimes referred to as the direct current (DC) conductivity, or simply the conductivity, in S/m, $\alpha \cong 0$ is an empirical parameter that describes the distribution of the relaxation times, and $\varepsilon_0$ denotes the dielectric constant in a vacuum (8.854·×10$^{-12}$ F/m). The simplicity of the Debye expression is deceptive, because $\varepsilon_{stat}$, $\tau$ and $\sigma_{stat}$ are all functions of the temperature $t$ and salinity $s$ of the sea water.

The expressions for $\varepsilon_W^{(C)}$, $\varepsilon_W$, and $\sigma_W$ as a function of water temperature $t$, salinity $s$, and the frequency $f$ of the electromagnetic wave propagation were developed by Klein and Swift (1977).

One point appears worth noting:

- If the water is analyzed in the HF band ($\omega_0 = 2\pi f_0$, $f_0 <1$ GHz), and is characterized by low salinity $s_{low}$ ($s_{low} \rightarrow 1$ ppt) for any temperature $t$, or by intermediate salinity $s_{low} < s < s_{up}$ ($s_{up} \approx 40$ ppt) only at high temperatures $t > t_{up}$ ($t_{up} \approx 29$ °C), then the complex relative dielectric permittivity of water $\varepsilon_W^{(C)}(t,s,\omega)$ can be approximated to the real dielectric permittivity $\varepsilon_W(t,s,\omega)$, thereby ignoring the electrical conductivity $\sigma_W(t,s,\omega)$. The relative dielectric permittivity of water $\varepsilon_W(t,s,\omega)$ can be approximated to its static value $\varepsilon_{stat}(t,s)$, even in the HF band (3 MHz to 30 MHz).

### 3.1. Relative dielectric permittivity of water and distance from the soil surface

The relative dielectric permittivity of the aqueous phase is lower than that of free water, because of interfacial solid–liquid forces. The dependence of this reduction on the moisture content and on the specific surface area is represented using a general approximated relationship by Friedman (1998). The model prediction is based on readily available soil properties (porosity, specific surface area, or texture), and it does not require any calibration.

As insufficient information is available on the real relaxation processes, and for the sake of generality, in the present study, the dielectric permittivity is assumed to grow exponentially $\varepsilon_W(z) = \varepsilon_W^{(low)} + (\varepsilon_W^{(up)} - \varepsilon_W^{(low)})(1-e^{-\lambda z})$, with minimum permittivity at infinite frequency $\varepsilon_W^{(low)} = \varepsilon_\infty \cong 4.9$ (Klein and Swift, 1977), and maximum permittivity at the value of 'free' water, i.e. static permittivity $\varepsilon_W^{(up)}(t,s) = \varepsilon_{stat}(t,s)$, at a film thickness $z$ of approximately two to three adsorbed water molecules, giving an averaged thickness of bonded water shell $d_{BW} = 1/\lambda$ varying in the range $\lambda = 10^7\text{-}10^9$ cm$^{-1}$.

The water shell thickness $d_W$ is calculated by dividing the volumetric content $\theta_W$ of the water contained in a mass unit $\rho_b$ of bulk soil by the specific surface area $S_{SA}$ of its solid phase, $d_W = \theta_W/(\rho_b \cdot S_{SA})$; similarly, the thickness of a bonded water shell $d_{BW}$ can be defined in terms of the volumetric content $\theta_{BW}$ of bonded water, $d_{BW} = \theta_{BW}/(\rho_b \cdot S_{SA})$, such that: $\theta_{BW} = (\rho_b \cdot S_{SA})/\lambda$. For terrestrial soils without a surface area measurement, $S_{SA}$ can be estimated from a given texture, according to the correlation of Banin and Amiel (1970), based on 33 Israeli soil samples of a wide range of textures: $S_{SA}$ (m$^2$/g) = 5.780 × % clay – 15.064. Thus, the averaged dielectric permittivity $\langle \varepsilon_W \rangle$ of the aqueous phase is represented by the harmonic mean of the local permittivity $\varepsilon_W(z)$ along the thickness $d_W$ of the water shell, i.e., $1/\langle \varepsilon_W \rangle = 1/\langle d_W \rangle \cdot \int_0^{d_W} dz/\varepsilon_W(z)$. Friedman (1998) solved the integral in a bluntly form, which is here rearranged more elegantly as:

$$\langle \varepsilon_W \rangle = \frac{\varepsilon_W^{(up)} \theta_W / \theta_{BW}}{\ln[1 + \frac{\varepsilon_W^{(up)}}{\varepsilon_W^{(low)}}(e^{\theta_W/\theta_{BW}} - 1)]} . \qquad (1)$$

## 4. Electrical conductivity, porosity and salinity

Using DC electrical conductivity values measured for a large number of brine-saturated core samples from a wide variety of sand formations, Archie (1942) described an empirical law:

$\sigma/\sigma_W = 1/F = a\, \eta^m$. Here $\sigma_W$ is the water conductivity, $F$ is the formation factor, $\eta$ is the porosity, and $m$ is the cementation index.

Subsequently, this Archie law has become an essential element in electric-log interpretation. The Archie law has been shown to hold true even for igneous rocks. However, clays can undergo ion exchange with a complicated conduction mechanism, and the Archie law does not hold for clayey rocks.

Sen et al. (1981) deliberately set out to define a model in which pore space was connected down to extremely low porosity values. They made a self-consistent approximation, which is known as the coherent potential approximation.

The De Loor (1964) theoretical model is conceived as a self-consistent formula for coated spheres, which avoids the issue of which material is host and which is impurity. In other words, the form is determined entirely by the model geometry. The De Loor model can be applied to concrete, to obtain the electrical conductivity (Fig. 1a):

$$\sigma(t,s,\omega) = \sigma_W(t,s,\omega) \frac{2\eta}{3-\eta}. \tag{2}$$

Thus, for $\eta < 1/2$, Eq. (2) implies that the concrete would be conductive. Eq. (2) gives $\sigma \propto \eta^m$ with $m = 1$ for low $\eta$.

Sen et al. (1981) followed a very simple intuitive method of incorporating the clustering effects in a single-site effective medium theory. This method has another positive feature. The self-consistent approximation (or coherent potential approximation; used at each step) provides very good results when the concentration of perturbation tends towards zero. Secondly, the geometrical model has a self-similarity that is often observed in terrestrial soils, i.e., the soil appears to be the same at any magnification.

The Sen model can be applied to soils to obtain their conductivity (Fig. 1b):

$$\sigma(t,s,\omega) = \sigma_W(t,s,\omega)\eta^{3/2}. \tag{3}$$

Eq. (3) shows an example of Archie-type behavior: $\sigma$ goes to zero as $\eta$ goes to zero; the exponent can be different from 3/2 for different shapes. A comparison between the DC result and the HF result implies that variations of $\sigma$ with frequency are not great. Therefore, the electrical conductivity $\sigma$ can be related to salinity $s$ using suitable theoretical models [Eqs. (2) or (3)] that are consistent with the predictions of the Archie empirical law (Fig. 1).

Next, the influence of salinity on the measurement of electrical conductivity is considered. The mathematical tool best suited to this purpose applies the so-called functions of sensitivity (Murray-Smith, 1987), which formalize the intuitive concept of sensitivity as the ratio between the percentage error of certain physical quantities (due to the variations in some parameters), and the percentage error of the same parameters.

The sensitivity function $S_s^\sigma$ of conductivity $\sigma$ relative to the salinity $s$ is defined as:

$$S_s^\sigma(t,s,\omega) = \frac{\partial \sigma(t,s,\omega)}{\partial s} \cdot \frac{s}{\sigma(t,s,\omega)}. \tag{4}$$

One point appears worth noting:

- The sensitivity $S_s^\sigma(t,s,\omega)$ for the electrical conductivity $\sigma$ of concrete or terrestrial soil relative to the salinity $s$ of water is almost uniform $S_s^\sigma \cong 1$ when the salinity $s$ tends towards low values, and so there is a linear variation of conductivity $\sigma$ with $s$, i.e. (Fig. 1):

$$S_s^\sigma(t,s,\omega) = \frac{\partial \sigma(t,s,\omega)}{\partial s} \cdot \frac{s}{\sigma(t,s,\omega)} \stackrel{\sigma \propto \sigma_W}{=} S_s^{\sigma_W}(t,s,\omega) = \frac{d\sigma_W(t,s,\omega)}{ds} \cdot \frac{s}{\sigma_W(t,s,\omega)} \stackrel{s \to 0}{\cong} 1. \quad (5)$$

## 5. Dielectric permittivity and volumetric water content

The results of Topp et al. (1980) demonstrate that the relative dielectric permittivity is strongly dependent on the volumetric content of water in terrestrial soil. In addition, dielectric permittivity is almost independent of soil density, texture, and salt content, and there is no significant temperature dependence.

A third-degree polynomial equation is fitted to the data from various mineral soils. The equation for this line is $\varepsilon(\theta_W) = 3.03 + 9.3\ \theta_W + 146.0\ (\theta_W)^2 - 76.7\ (\theta_W)^3$. This equation is constrained to pass through (81.5, 1), which is the data point for pure water at 20 °C.

In practice, the permittivity $\varepsilon$ is usually measured and the water content $\theta_W$ is determined. The following equation assumes $\varepsilon$ is known and $\theta_W$ is found (Fig. 2):

$$\theta_W(\varepsilon) = -5.3 \times 10^{-2} + 2.92 \times 10^{-2}\varepsilon - 5.5 \times 10^{-4}\varepsilon^2 + 4.3 \times 10^{-6}\varepsilon^4. \quad (6)$$

### 5.1. Theoretical models for permittivity

Many types of dielectric models have been developed to satisfy a wide range of circumstances, and there are a number of comprehensive reviews of the topic in the literature. For the present study, these can be broadly classified into geometric models and simple volumetric dielectric models (Tsui and Matthews, 1997; Friedman, 1998).

Geometric dielectric models are used in an effort to provide a representation of the physical nature of the mixture in question. These models offer a greater range of applicability than simple volumetric models, and they represent much more complicated formulations, with the associated difficulties to achieve numerical solutions, particularly when they address the effective complex dielectric permittivity $\varepsilon^{(C)}$ of mixtures that contain water.

The application of the Loor (1968) three-phase model assumes that material solids act as host materials, while treating the air and saline components as spherical inclusions in the host material forming the mixture. This model can be expressed mathematically as:

$$\frac{\varepsilon^{(C)} - \varepsilon_S}{3\varepsilon^{(C)}} = (\eta - \theta_W) \frac{\varepsilon_A - \varepsilon_S}{\varepsilon_A + 2\varepsilon^{(C)}} + \theta_W \frac{\varepsilon_W^{(C)} - \varepsilon_S}{\varepsilon_W^{(C)} + 2\varepsilon^{(C)}}, \qquad (7)$$

where $\varepsilon_W^{(C)}$ is the complex relative dielectric permittivity of the water phase (Klein and Swift, 1977); $\varepsilon_A = 1.0$ and $\varepsilon_S$ are the relative dielectric permittivities of the air and solid phases, respectively; $\eta$ is the porosity and $\theta_W$ is the volumetric water content.

A volumetric model considers only the volume fraction of the components. A large number of different formulae exist for the effective complex dielectric permittivity $\varepsilon^{(C)}$ of mixtures, and these are often used without ascertaining whether the sample conforms to the geometry for which the formula holds in each specific case. The derivation assumes a model of parallel layers with layer thicknesses much greater than the wavelength.

The Complex Refractive Index Model (CRIM) asserts that the effective complex refractive index for a mixture is provided by the volumetric mean of the refractive indices of the components (Robinson et al., 2002):

$$\sqrt{\varepsilon^{(C)}} = (1-\eta)\sqrt{\varepsilon_S} + \theta_W \sqrt{\varepsilon_W^{(C)}} + (\eta - \theta_W)\sqrt{\varepsilon_A}. \qquad (8)$$

The CRIM model has been widely used for terrestrial soil varieties due to its simplicity; however, this method is not applicable for calculating the relative dielectric permittivity of concrete. The reason for this is because the CRIM model is a function of the volume fraction, but it does not consider the geometrical shape and orientation of inclusions. It is generally considered to be inaccurate in contexts of high salinity or LF. Sometimes, the de Loor and CRIM laws are applied as real mixture laws to predict the real part of dielectric permittivity $\varepsilon$ by considering only the real part $\varepsilon_W$ and ignoring the imaginary part $\sigma_W/\omega\varepsilon_0 \ll \varepsilon_W$ of the water permittivity.

In the Appendix, for both concrete and terrestrial soil, it is underlined that the relative dielectric permittivity $\varepsilon$ is related to the volumetric water content $\theta_W$ by way of suitably refined theoretical models that are consistent with the predictions of the Topp empirical law. The better the agreement, the lower the hygroscopic water content $\theta_W$ and the higher the dielectric permittivity $\varepsilon$ (Fig. 2); consequently the best agreement is achieved with concrete containing almost no bonded water molecules, and only if characterized by high electrical conductivity.

## 6. The RESPER probe

In previous studies (Settimi et al., 2010a b) and in a recent report (Settimi et al., 2011) the authors proposed a discussion of theoretical modeling and a move towards the practical implementation of an induction probe that can acquire transfer impedance in the field.

A RESPER probe enables measurements of electrical resistivity and dielectric permittivity using alternating currents in LF (30 kHz $<f<$ 300 kHz) and MF (300 kHz $<f<$ 3 MHz), and up to HF (3 MHz $<f<$ 30 MHz) bands. The measurements are taken using four electrodes laid on the surface to be analyzed, and through measurements of complex impedance, the resistivity and permittivity of the material can be established. Furthermore, by increasing the distance between the electrodes, the electrical properties of the sub-surface structures can be investigated to greater depths. The main advantage of the RESPER probe is that measurements of electrical parameters can be conducted in a nondestructive manner, thereby enabling characterization of precious and unique materials. Also, in appropriate arrangements, measurements can be taken with the electrodes raised slightly above the surface, providing totally noninvasive analysis, although this is accompanied by greater errors. The RESPER probe can perform measurements on materials with high resistivity and permittivity in an immediate way, without the need for later stages of data post-analysis.

An initial study (Settimi et al., 2010a) discussed the theoretical modeling of an induction probe that performs simultaneous noninvasive measurements of electrical resistivity $1/\sigma$ and dielectric permittivity $\varepsilon$ of non-saturated media (using a RESPER probe). A mathematical–physical model was applied on the propagation of errors in the measurements of resistivity and permittivity, based on the sensitivity functions tool. The findings were also compared with the results of the classical method of analysis in the frequency domain, which is useful for determining the behavior of zero and pole frequencies in the linear time invariant circuit of the RESPER probe. The study underlined that mean values of electrical resistivity and dielectric permittivity can be used to estimate the complex impedance over various concrete and terrestrial soil types, especially when they are characterized by low volumetric water content and analyzed within a LF bandwidth. To meet the design specifications required to ensure satisfactory performance of the RESPER probe, the forecasts of the sensitivity-functions approach are more reliable than the results foreseen by the transfer-functions method. In other words, once the measurement inaccuracy is within an acceptable limit (10%*)*, the sensitivity approach provides more realistic values, as compared to those provided by the transfer method. These numeric values concern both the band of frequency $f$ for the probe and the measurable range of resistivity $1/\sigma$ or permittivity $\varepsilon$ for the concrete and soil (the order of magnitude of these values is reported in the relevant literature; see Settimi et al., 2010a, and references therein).

A second study (Settimi et al., 2010b) moved towards practical implementation of electrical spectroscopy. To design a RESPER probe to perform measurements of *1/σ* and *ε* on a subsurface with inaccuracies below a prefixed limit (10%) in a bandwidth of MFs, the RESPER probe should be connected to an appropriate ADC that can sample in phase and quadrature (IQ), or in uniform mode. If the probe is characterized by galvanic contact with the surface, then the inaccuracies in the measurement of resistivity and permittivity due to the IQ or uniform sampling ADC can be expressed analytically. A large number of numerical simulations have shown that performance depends on the selected sampler, and that under the same operating conditions, the IQ is better, as compared to the uniform mode; i.e. number of bits and medium (see Settimi et al., 2010b, and references therein).

The analysis showed that the RESPER probe can work at an optimum MF if the transfer impedance is characterized by a MF cut-off frequency, which is in agreement with more traditional results in the literature (Grard, 1990a, b; Grard and Tabbagh, 1991; Tabbagh et al., 1993; Vannaroni et al. 2004; Del Vento and Vannaroni, 2005). Unlike these previous studies, the probe can perform measurements up to an appropriate band of higher frequencies than the cut-off frequency, where the inaccuracy for the measurements of conductivity and permittivity remain below the fixed limit.

Finally, a recent report (Settimi et al., 2011) discussed the preliminary design of a RESPER probe prototype, moving towards its configuration in a multi-dipole-dipole array (for further technical information and the data sheet, the reader is referred to Settimi et al., 2011).

## 6.1. IQ sampling ADC

Let us consider the IQ mode (Jankovic and Öhman, 2001). The IQ quartz is oscillating with a period $T$ that is affected by an inaccuracy $\Delta T/T$. The quartz figure of merit $Q = T/\Delta T$ assumes high enough values, i.e. $1/Q << 1/(2\pi)$ $[Q=10^4\text{-}10^6]$. In the limit case, corresponding to $Q \rightarrow \infty$, it can be shown that the complex impedance $Z$ can be measured with a modulus inaccuracy $\Delta|Z|/|Z|(n)$ that depends on the bit resolution $n$, decreasing as the power function $2^{-n}$ of $n$; i.e. (Settimi et al., 2010b):

$$\frac{\Delta|Z|}{|Z|} = \frac{1}{2^n}, \qquad (9)$$

while the phase inaccuracy $\Delta\Phi_Z/\Phi_Z(n,\varphi_V)$ depends both on the resolution $n$, still decreasing as the power function $2^{-n}$ of $n$, and on the voltage phase $\varphi_V$, such that (Settimi et al., 2010b):

$$\frac{\Delta\Phi_Z}{\Phi_Z} = \frac{1}{2^n}\frac{\sin(2\varphi_V)}{2\varphi_V} = \begin{cases} \frac{1}{2^n} &, \varphi_V = \varphi_V^{\max} = 0 \\ 0 &, \varphi_V = \varphi_V^{\min} = \pi \end{cases}. \qquad (10)$$

With the aim of investigating the physics of the measuring system, if the RESPER probe shows galvanic contact with the subjacent medium of electrical conductivity $\sigma$ and dielectric permittivity $\varepsilon$, and works at frequencies $\omega=2\pi f$ lower than the cut-off frequency $\omega_T=\omega_T(\sigma,\varepsilon_r)=\sigma/(\varepsilon_0(\varepsilon_r+1))$ (Settimi et al., 2010a),

$$\Omega = \frac{\omega}{\omega_T} \leq 1, \qquad (11)$$

then the inaccuracies $\Delta\sigma/\sigma$ in the measurements of conductivity $\sigma$ and $\Delta\varepsilon/\varepsilon$ for permittivity $\varepsilon$ are expressed analytically, when the RESPER probe is connected to a IQ sampler which ensures the inaccuracies $\Delta|Z|/|Z|$ (9) for modulus $|Z|$ and $\Delta\Phi_Z/\Phi_Z$ (10) for phase $\Phi_Z$ of the complex impedance (Settimi et al., 2010b),

$$\frac{\Delta\sigma}{\sigma} \cong (1+\Omega^2)(\frac{\Delta|Z|}{|Z|} + \frac{\Delta\Phi_Z}{\Phi_Z}), \qquad (12)$$

$$\frac{\Delta\varepsilon}{\varepsilon} \cong (1+\Omega^2)(1+\frac{1}{\varepsilon})(\frac{1}{\Omega^2}\frac{\Delta|Z|}{|Z|} + \frac{\Delta\Phi_Z}{\Phi_Z}). \qquad (13)$$

Only if the RESPER is in galvanic contact with the medium does the mathematical–physical model predict that the inaccuracies $\Delta\sigma/\sigma$ for $\sigma$ and $\Delta\varepsilon/\varepsilon$ for $\varepsilon$ are invariant in the multi-dipole-dipole configuration, and independent of the characteristic geometrical dimension of the probe, i.e. the electrode–electrode distance (Settimi et al., 2011). If besides grazing the medium, the RESPER measures $\sigma$ and $\varepsilon_r$ working in a frequency $\omega$ that is much lower than the cut-off frequency $\omega_T=\omega_T(\sigma,\varepsilon_r)$, then the inaccuracy $\Delta\sigma/\sigma=F(\Delta|Z|/|Z|,\Delta\Phi_Z/\Phi_Z)$ is a linear combination of the inaccuracies, $\Delta|Z|/|Z|$ and $\Delta\Phi_Z/\Phi_Z$, for complex impedance, while the inaccuracy $\Delta\varepsilon/\varepsilon=F(\Delta|Z|/|Z|)$ can be approximated as a linear function only of the inaccuracy $\Delta|Z|/|Z|$; in other words, if $\omega<<\omega_T$, then $\Delta\Phi_Z/\Phi_Z$ contributes to $\Delta\sigma/\sigma$ but not to $\Delta\varepsilon/\varepsilon$ (Fig. 3).

Referring to the IQ sampling ADC, the inaccuracies $\Delta\sigma/\sigma$ and $\Delta\varepsilon/\varepsilon$ are estimated for the worst case. So, the inaccuracies $\Delta|Z|/|Z|(n)$ and $\Delta\Phi_Z/\Phi_Z(n,\varphi_V)$ assume the mean and the maximum values, respectively; i.e. $\Delta|Z|/|Z| = \Delta\Phi_Z/\Phi_Z = 1/2^n$.

One point appears worth noting:

- Within the limit of the HFs (3 MHz to 30 MHz), and always satisfying the condition:

$$\sigma/\omega\varepsilon_0 \ll 1, \tag{14}$$

the inaccuracy $\Delta\varepsilon/\varepsilon$ of the relative dielectric permittivity $\varepsilon$ measurements is minimized to the value:

$$\varepsilon^{(min)} \cong \frac{1}{2}[1+\frac{1}{3}(\frac{2\sigma}{\omega\varepsilon_0})^2], \tag{15}$$

which is a quadratic function that decreases with frequency $\omega$ and increases with electrical conductivity $\sigma$, but is not dependent on porosity η of either the concrete or the terrestrial soil (Fig. 3).

## 7. Salinity and water content inaccuracies of the RESPER

A novelty of the present report is the application of a mathematical–physical model to the propagation of errors in the measurements based on a sensitivity functions tool (Murray-Smith, 1987). The inaccuracy $\Delta s/s$ of salinity is the ratio between the inaccuracy $\Delta\sigma/\sigma$ of electrical conductivity, specified by the RESPER probe [Eqs. (9), (10) and (12)], and the function of sensitivity $S_s^\sigma$ for conductivity relative to salinity, derived using the constitutive equations of the medium [Eqs. (2), (3) and (4)], i.e.,

$$\frac{\Delta s}{s}(t,s,\omega,\sigma,\varepsilon) = \frac{1}{S_s^\sigma(t,s,\omega)} \frac{\Delta\sigma}{\sigma}(\omega,\sigma,\varepsilon). \qquad (16)$$

Instead, the inaccuracy $\Delta\theta_W/\theta_W$ for volumetric content of water is the product of the inaccuracy $\Delta\varepsilon/\varepsilon$ of relative dielectric permittivity, specified by the RESPER [Eqs. (9), (10) and (13)], and the sensitivity function $S_\varepsilon^{\theta_W}$ of volumetric water content relative to dielectric permittivity, derived from the constitutive equations of the medium [Eqs. (19)-(21) and (28)-(30)], i.e.,

$$\frac{\Delta\theta_W}{\theta_W}(t,s,\omega,\sigma,\varepsilon) = S_\varepsilon^{\theta_W}(t,s,\omega,\varepsilon) \frac{\Delta\varepsilon}{\varepsilon}(\omega,\sigma,\varepsilon). \qquad (17)$$

The main result is the model prediction that according to Eqs. (16) and (17), the lower the inaccuracy for the measurements of $s$ and $\theta_W$ (decreasing by as much as one order of magnitude, from 10% to 1%) the higher the $\sigma$; so that inaccuracy for terrestrial soil is lower (Figs. 4 and 5). The proposed physical explanation is that water molecules are mostly dispersed as $H^+$ and $OH^-$ ions throughout the volume of concrete, but are almost all concentrated as bonded $H_2O$ molecules only at the surface of soil.

The following point is worth noting:
- The inaccuracy $\Delta\theta_W/\theta_W$ [Eq. (17)] in measurements of volumetric water content $\theta_W(\varepsilon)$ performed using a RESPER probe diverges [Eq. (23)] and is minimized [Eq. (24)] or maximized [Eqs. (32) and (33)] into values that are not dependent on the bit resolution $n$ [Eqs. (9) and (10)] of the IQ ADC.

## 8. Conclusions

This report has discussed the performance of electrical spectroscopy using a RESPER probe to measure salinity $s$ and volumetric water content $\theta_W$ of concrete and terrestrial soil. The RESPER probe is an induction device for spectroscopy that performs simultaneous noninvasive measurements of the electrical resistivity $1/\sigma$ and relative dielectric permittivity $\varepsilon_r$ of a subjacent medium. Numerical simulations have established that the RESPER can measure $\sigma$ and $\varepsilon$ with inaccuracies below a predefined limit (10%) up to the HF band. Conductivity is related to salinity [Eqs. (2) or (3), and Fig. 1] and dielectric permittivity to volumetric water content [Eqs. (18) or (26), (27)] using suitably refined theoretical models that are consistent with the predictions of the Archie and Topp empirical laws. The better agreement, the lower the hygroscopic water content and the higher the $s$ (Fig. 2); so there is closer agreement with concrete containing almost no bonded water molecules, provided these are characterized by a high $\sigma$. A novelty of the present report is the application of a mathematical–physical model to the propagation of errors in the measurements based on a sensitivity functions tool. The inaccuracy of salinity [Eq. (16)] (water content [Eq. (17)]) is the ratio (product) between the conductivity (permittivity) inaccuracy, specified by the probe [Eqs. (9)-(13), and Fig. 3], and the sensitivity function of salinity (water content) relative to conductivity [Eqs. (4) and (5)] (permittivity [Eqs. (19)-(21) and (28)-(30)]), derived from the constitutive equations of the medium. The main result is the model prediction that the lower the inaccuracy for the measurements of $s$ and $\theta_W$ (decreasing by as much as an order of magnitude from 10% to 1%), the higher the $\sigma$ (Figs. 4 and 5); so the inaccuracy for soils is lower (Table 1). The proposed physical explanation is that water molecules are mostly dispersed as $H^+$ and $OH^-$ ions throughout the volume of concrete, but are almost all concentrated as bonded $H_2O$ molecules only at the surface of soil.

The following points are worth noting:

- If the water analyzed in the HF band ($\omega_0 = 2\pi f_0$, $f_0 <1$ GHz) is characterized by low salinity $s_{low}$ ($s_{low} \to 1$ ppt) for any temperature $t$ or by intermediate salinity $s_{low} < s < s_{up}$ ($s_{up} \approx 40$ ppt) but only at high temperatures $t > t_{up}$ ($t_{up} \approx 29$ °C), then the complex relative dielectric permittivity of water $\varepsilon_W^{(C)}(t,s,\omega)$ can be approximated to its real part $\varepsilon_W(t,s,\omega)$, thereby ignoring the electrical conductivity $\sigma(t,s,\omega)$.

  The relative dielectric permittivity of water $\varepsilon_W(t,s,\omega)$ can be approximated to its static value $\varepsilon_{stat}(t,s)$ even in the HF band (3 MHz to 30 MHz) [see (Klein and Swift, 1977)].

- When the water phase analyzed in the HF band is characterized by low salinity, the temperature $t$ has almost no influence on the measurements of the relative dielectric permittivity values $\varepsilon_W(t,s,\omega)$ for water, and $\varepsilon(t,s,\omega)$ for concrete and terrestrial soil, and so on their volumetric water content $\theta_W(\varepsilon)$.

  For each non-saturated material variety ($\alpha \to 0$), the frequency $\omega = 2\pi f$ influences the salinity $s(\sigma)$ measurements but not the volumetric water content $\theta_W(\varepsilon)$ measurements [Fig. 2], because even for HFs, water is characterized by an electrical conductivity $\sigma(t,s,\omega)$ that varies quadratically with $\omega$,

$$\sigma_W(t,s,\omega) \cong \sigma_{stat}(t,s) + \omega^2 \varepsilon_0 \tau [\varepsilon_{stat}(t,s) - \varepsilon_\infty]$$

and a dielectric permittivity $\varepsilon_W(t,s,\omega)$ that remains constant with $\omega$,

$$\varepsilon_W(t,s,\omega) \cong \varepsilon_{stat}(t,s),$$

where $\varepsilon_\infty$ is the permittivity at infinite frequency, $\varepsilon_{stat}$ is the static permittivity, $\tau$ is the relaxation time in s, $\sigma_{stat}$ is the ionic or ohmic conductivity, which is sometimes referred to as the DC conductivity or simply the conductivity, in S/m, and $\varepsilon_0$ denotes the dielectric constant in a vacuum ($8.854 \cdot \times 10^{-12}$ F/m) [see (Klein and Swift, 1977)].

Furthermore, the mathematical–physical model describing the dielectric properties of concrete shows that the volumetric water content $\theta_W(t,s,\omega,\varepsilon)$, which is a function of the relative dielectric permittivity $\varepsilon$, shows almost no dependence on frequency $\omega$, salinity $s$ and temperature $t$ because the dielectric permittivity value $\varepsilon$ is much lower than water permittivity, $\varepsilon << \varepsilon_W(t,s,\omega)$ [Eq. (18), and Fig. 2a].

- The function of sensitivity $S_s^\sigma(t,s,\omega)$ for the electrical conductivity $\sigma$ of concrete and terrestrial soil relative to the salinity $s$ of water is almost uniform $S_s^\sigma \cong 1$ when the salinity $s$ tends towards low values, and so there is a linear variation of conductivity $\sigma$ with $s$ [Eqs. (4), (5); Figs. 1 and 4].

- The sensitivity function $S_\varepsilon^{\theta_W}(t,s,\omega,\varepsilon)$ for the volumetric content $\theta_W$ of water relative to the dielectric permittivity $\varepsilon$ shows that:

With reference to concrete, it depends on frequency $\omega$ to a small extent, but has almost no dependence on either temperature $t$ or salinity $s$, especially as under operating conditions such that $\sigma_W/\omega\varepsilon_0 << \varepsilon_W$; it diverges to infinity and is minimized into the values $\varepsilon^{(asym)}$ and $\varepsilon^{(knee)}$, respectively, which are functions of the porosity η for concrete, and of both the dielectric permittivity values $\varepsilon_A$ and $\varepsilon_S$ for air and the solid components; furthermore, the value $\varepsilon^{(knee)}$ also depends on water permittivity $\varepsilon_W$ [Eqs. (19)-(24), and Fig. 5a];

With reference to terrestrial soil, its sensitivity has minimal dependence on frequency $\omega$, especially under operating conditions such that $\sigma_W/\omega\varepsilon_0 << \varepsilon_W$; and it is maximized into the refractive index $\sqrt{\varepsilon^{(max)}}$, which is a linear combination of the refractive indices $\sqrt{\varepsilon_A}$ and $\sqrt{\varepsilon_S}$ for air and the solid components of soil, respectively, and it depends on their porosity η [Eqs. (26), (28)-(34), and Fig. 5b].

- Within the limit of HFs, the inaccuracy $\Delta\varepsilon/\varepsilon$ of the relative dielectric permittivity $\varepsilon$ measurement is minimized into the value $\varepsilon^{(min)}$, which is a quadratic function that decreases with frequency $\omega$ and increases with electrical conductivity $\sigma$, but is not dependent on the porosity η of either the concrete or terrestrial soil [Eqs. (14), (15), and Fig. 3].

- The inaccuracy $\Delta\theta_W/\theta_W$ in measurements of volumetric water content $\theta_W(\varepsilon)$ performed by using a RESPER probe diverges and is minimized or maximized into values that are not dependent on the bit resolution $n$ of the IQ ADC [Eqs. (9)-(13) and (16), (17)].

**Acknowledgments.**

The author would like to thank Prof. G. Santarato for interesting discussions on the measurement of salinity and water content, and also for useful pointers for literature regarding concrete and terrestrial soil. Moreover, the author is very grateful to Prof. S. Friedman for an academic discussion useful to shed light on the derivation of Equation (1).

**References.**

**Table 1.** Operating conditions described in the captions of Figures 1 and 2. The theory: modeling water content $\theta_W(t_{up}, s_{low}, f_0, \varepsilon)$ as a function of relative dielectric permittivity $\varepsilon$ is always valid for all the concrete samples and holds up to a reasonable limit $\theta_{W,lim}(t_{up}, s_{low}, f_0)$ for fine or coarse textured terrestrial soil varieties, with low or high electrical resistivity.

| Soil | Low Resistivity | High Resistivity |
|---|---|---|
| Fine Textured ($\rho_b = 1.2$ g/cm$^3$) | $\theta_{W,lim}(t_{up}, s_{low}, f_0) = 0.219$ | $\theta_{W,lim}(t_{up}, s_{low}, f_0) = 2.191 \cdot \times 10^{-3}$ |
| Coarse Textured ($\rho_b = 1.6$ g/cm$^3$) | $\theta_{W,lim}(t_{up}, s_{low}, f_0) = 0.292$ | $\theta_{W,lim}(t_{up}, s_{low}, f_0) = 2.922 \cdot \times 10^{-3}$ |

**Figure captions.**

**Figure 1.** A material medium analyzed at low and high temperatures ($t_{low}$ = -2 ºC, $t_{up}$ = 29 ºC) and within the HF band ($f_0$ < 1 GHz). The medium can be concrete (a) or terrestrial soil (b). Concrete (Cheeseman et al., 1998): a three component mixture of water (Klein and Swift, 1977), air (relative dielectric permittivity, $\varepsilon_A$ = 1), and solid (ordinary Portland cement) phases; fine or coarse textured, with a high or low water to cement ratio, respectively ($W/C$ = 0.4-0.5); and high or low electrical resistivity, respectively, with dielectric permittivity $\varepsilon_S$ = 4-7. Soil: a three component mixture of water, air, and solid (mineral $\varepsilon_S$ = 3.9 or organic $\varepsilon_S$ = 5) phases, with low or high thickness of the bonded water shell [$d_{BW}(\lambda)=1/\lambda$, $\lambda$ = $10^7$-$10^9$ cm$^{-1}$]; fine or coarse textured, respectively, with low or high loose bulk density ($\rho_b$ = 1.2-1.6 g/cm$^3$); and high or low resistivity, respectively, composed of pure clay minerals (apparent particle density, $\rho_p$ = 2.65 g/cm$^3$) or even organic matter (OM = 10%). The Sen et al. (1981) and De Loor (1964) theoretical models overlap well with the Archie (1942) empirical law. Plots of the electrical conductivity $\sigma(t_{low,up}, s, f_0)$, in units of S/m, as a function of the salinity $s$, in the range $s \in [s_{low}, s_{up}]$, with $s_{low}$ = 1 ppt and $s_{up}$ = 40 ppt, for both the concrete and soil.

**Figure 2.** Operating conditions are described in the caption of Figure 1. Concrete (a) and terrestrial soil (b) are characterized by high temperature ($t_{up}$ = 29 ºC) and low salinity ($s_{low}$ = 1 ppt). Overlap of the present theoretical model with the Topp et al. (1980) empirical law. Semi-logarithmic plots for the volumetric content $\theta_W(t_{up}, s_{low}, f_0, \varepsilon)$ of water as a function of relative dielectric permittivity $\varepsilon$, for both the concrete and soil (Table 1).

**Figure 3.** RESPER probe characterized by galvanic contact with a subjacent medium. The RESPER is connected to an ADC which samples in IQ mode and is specified by a minimum

bit resolution $n_{min}$ = 12, ensuring measurement inaccuracies below a predefined limit (10%) up to the HF band (Settimi et al., 2010a, b). The medium can be a variety of concrete or terrestrial soil. Concrete: with low or high electrical resistivity, and respectively high or low relative dielectric permittivity, i.e. ($1/\sigma$ = 4000 Ω·m, $\varepsilon$ = 9) or ($1/\sigma$ = 10000 Ω·m, $\varepsilon$ = 4). Soil: with low or high electrical resistivity, and respectively high or low dielectric permittivity, i.e. ($1/\sigma$ = 130 Ω·m, $\varepsilon$ = 13) or ($1/\sigma$ = 3000 Ω·m, $\varepsilon$ = 4). The probe performs measurements at HFs, and the media is analyzed at frequency $f_{low}$ = 3 MHz, apart from soils with low resistivity ($f_{up}$ = 30 MHz). The Like-Bode diagrams of inaccuracy $\Delta\sigma/\sigma(f_{low,up}, \sigma, \varepsilon)$ as a function of $\sigma$ (a) and semi-logarithmic plots of inaccuracy $\Delta\varepsilon/\varepsilon(f_{low,up}, \sigma, \varepsilon)$ as a function of permittivity $\varepsilon$ (b), for both the concrete and soil.

**Figure 4.** Operating conditions are described in the captions of Figures 1 and 3. Semi-logarithmic plots of the inaccuracy $\Delta s/s(t_{low,up}, s, f_{low,up}, \sigma)$ as a function of salinity $s$, in the range $s \in [s_{low}, s_{up}]$, with $s_{low}$ = 1 ppt and $s_{up}$ = 40 ppt, for both concrete (a) and terrestrial soil (b).

**Figure 5.** Operating conditions are described in the captions of Figures 1 and 3. The Like-Bode diagrams of inaccuracy $\Delta\theta_W/\theta_W (t_{up}, s_{low}, f_{low,up}, \theta_W)$ as a function of volumetric water content $\theta_W$, valid within the range $\theta_W \in [0, \theta_{W,\lim}]$ defined in Table 1, for both concrete (a) and soil (b).

**Figure 1a**

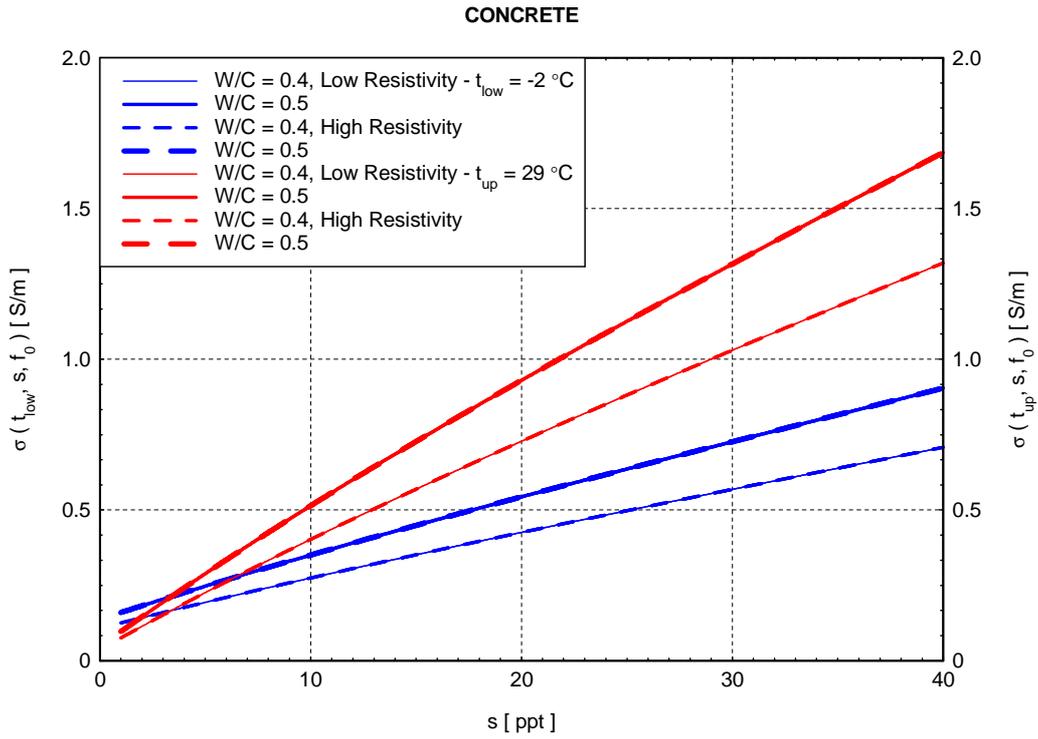

**Figure 1b**

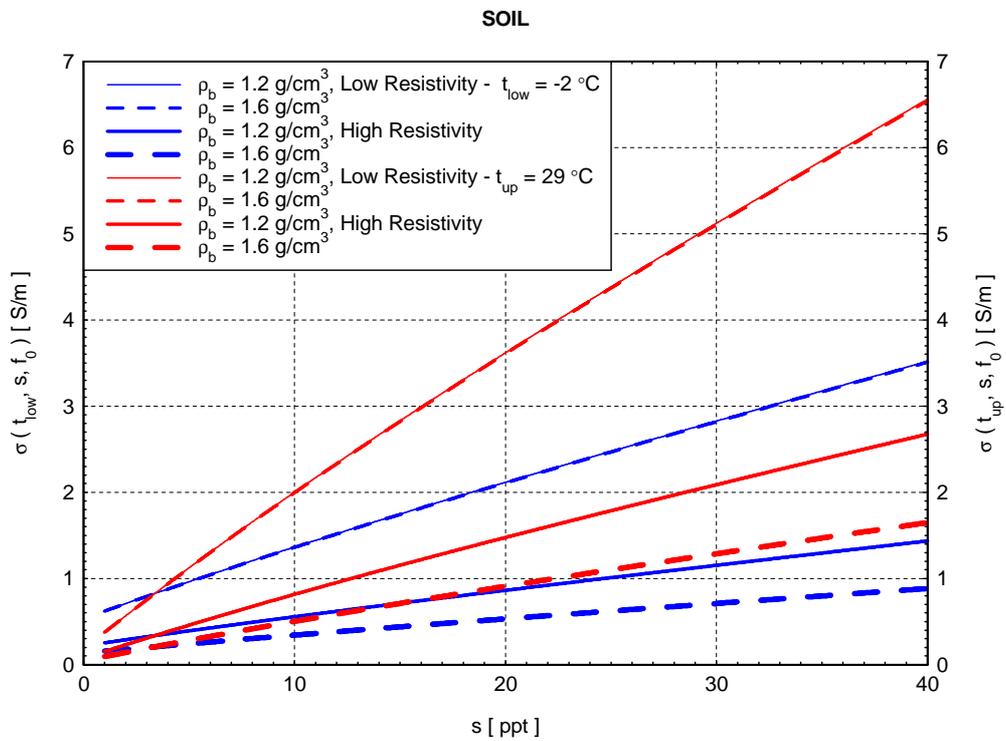

**Figure 2a**

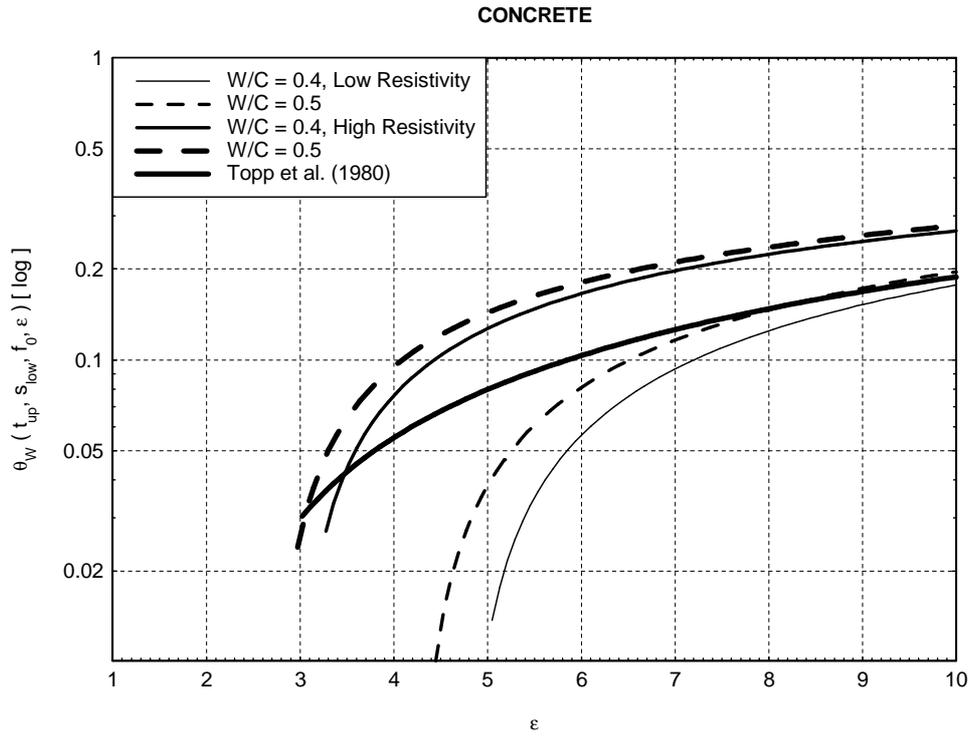

**Figure 2b**

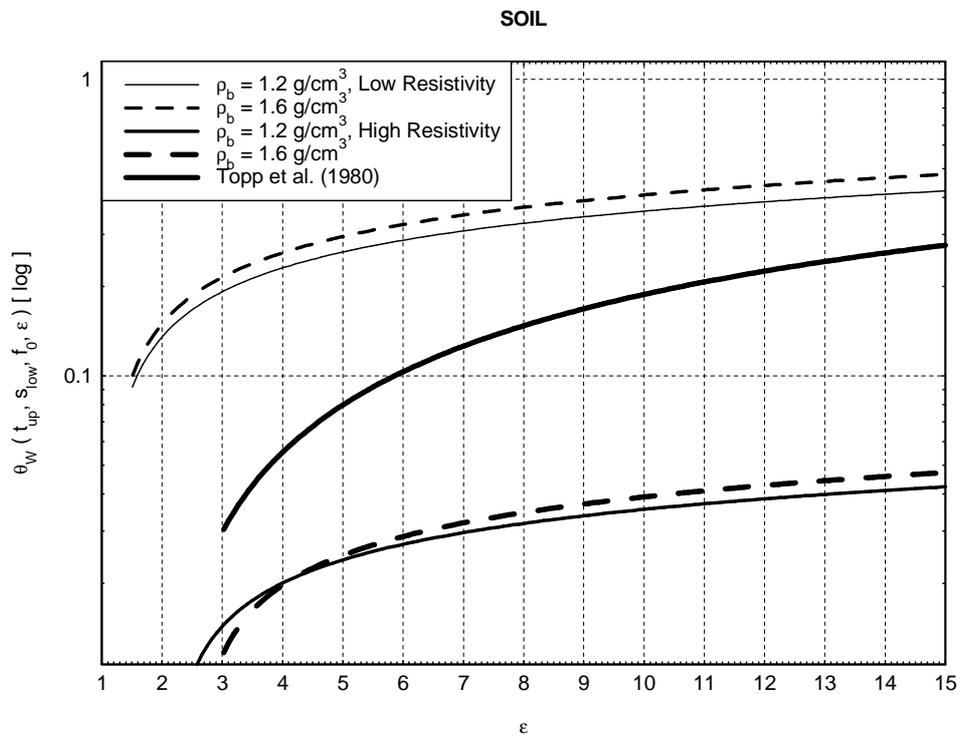

**Figure 3a**

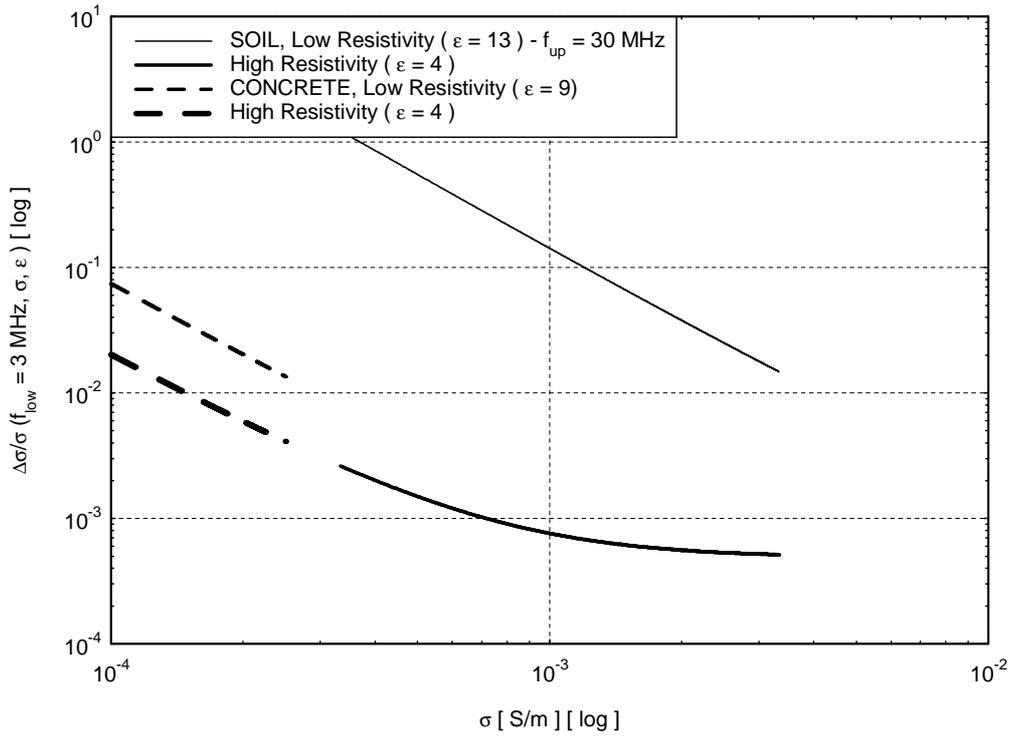

**Figure 3b**

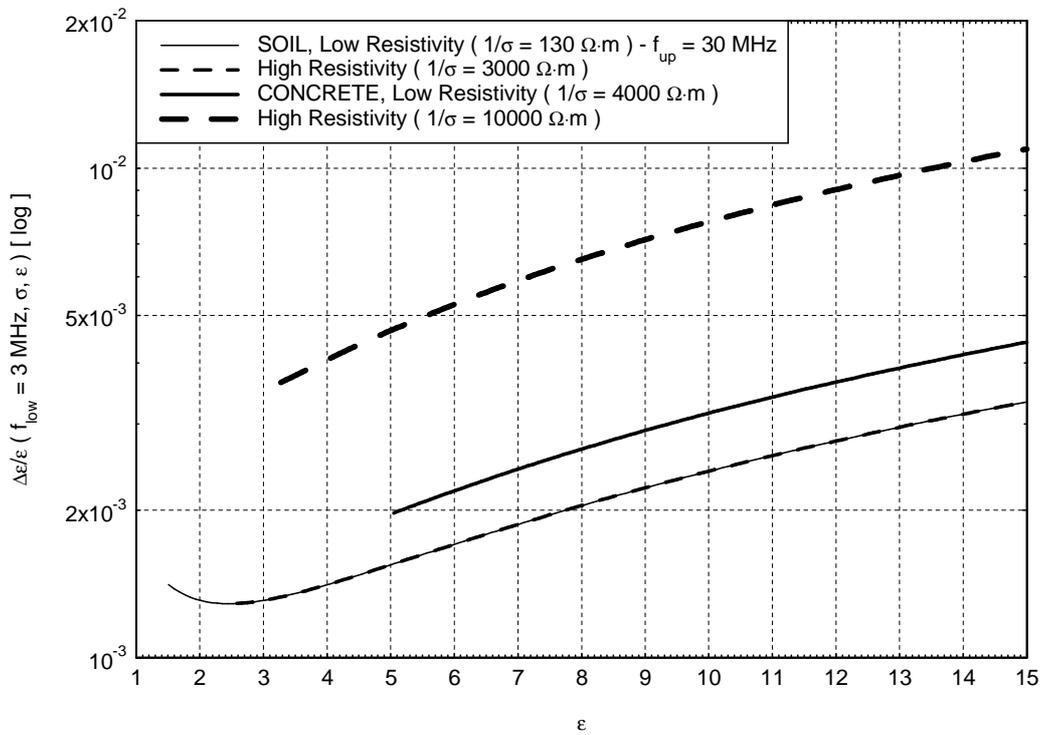

**Figure 4a**

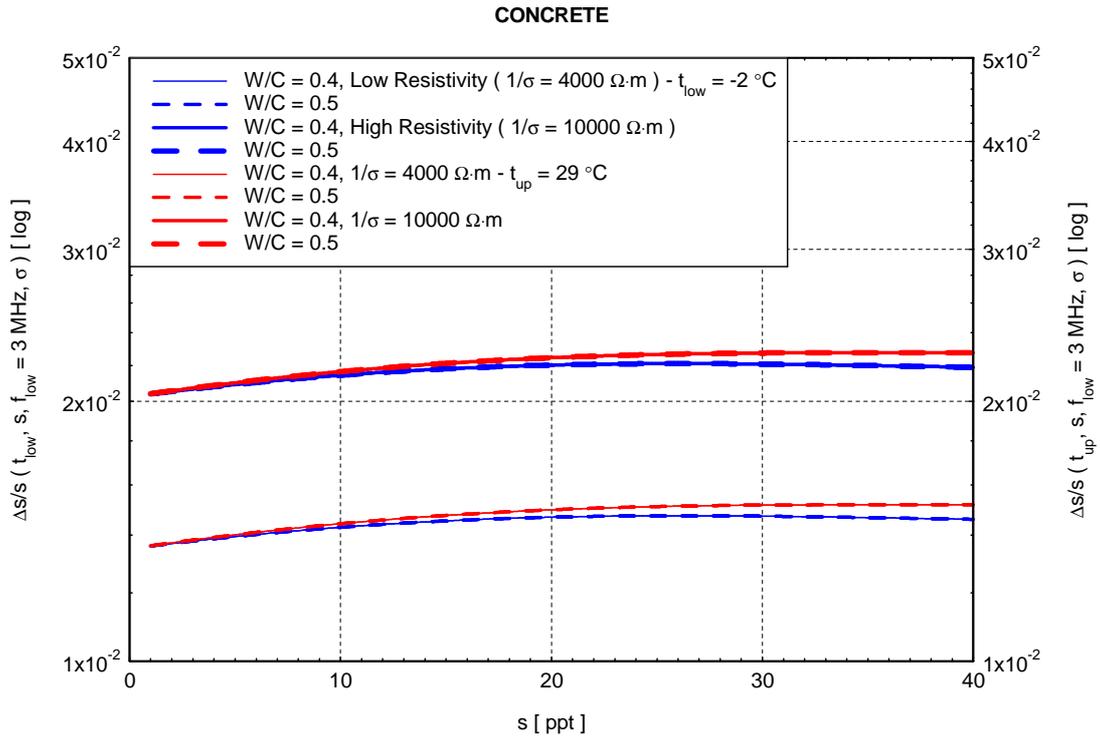

**Figure 4b**

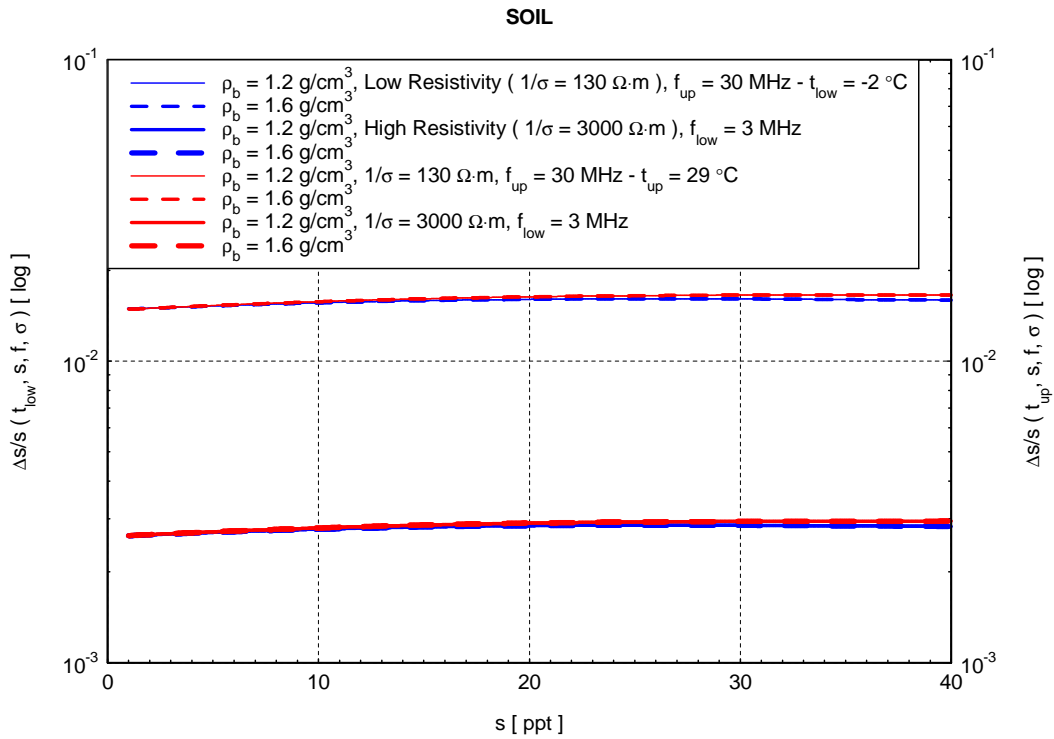

**Figure 5a**

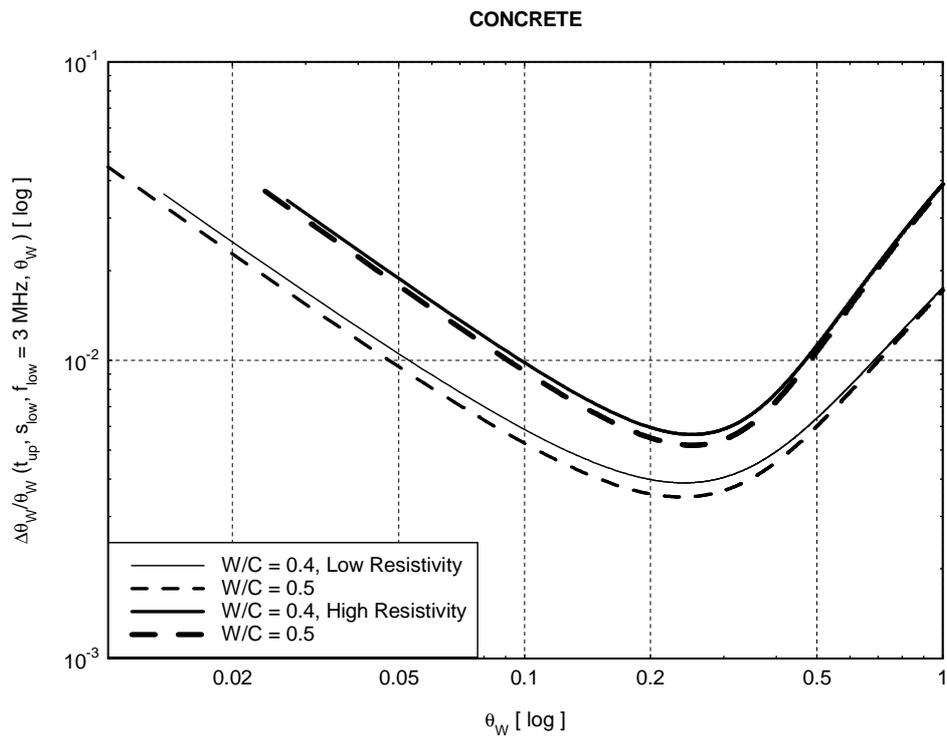

**Figure 5b**

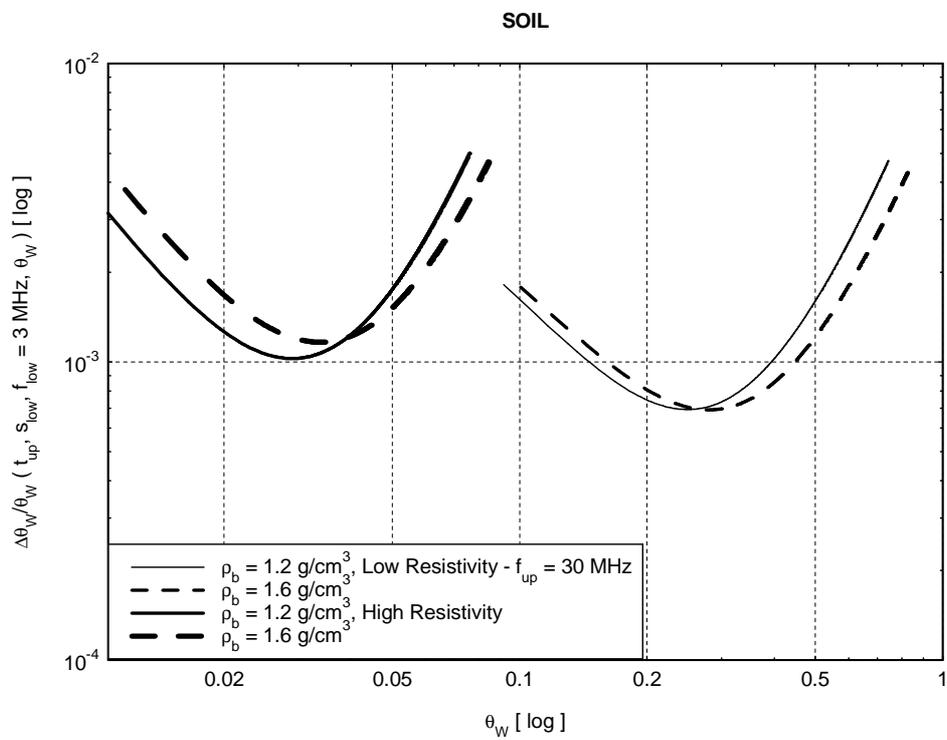

**Appendix.**

*Concrete.* For the hypothesis that $\sigma_W/\omega\varepsilon_0 \ll \varepsilon_W$, reversing Eq. (7), the volumetric content $\theta_W$ of water can be expressed as a function of the relative dielectric permittivity $\varepsilon$ (Fig. 2a):

$$\theta_W(t,s,\omega,\varepsilon) = \frac{\dfrac{\varepsilon-\varepsilon_S}{3\varepsilon} - \eta\dfrac{\varepsilon_A-\varepsilon_S}{\varepsilon_A+2\varepsilon}}{\dfrac{\varepsilon_W(t,s,\omega)-\varepsilon_S}{\varepsilon_W(t,s,\omega)+2\varepsilon} - \dfrac{\varepsilon_A-\varepsilon_S}{\varepsilon_A+2\varepsilon}}. \tag{18}$$

In all types of concrete, the most significant solid components are coarse aggregates, fine aggregates, and cement paste. Coarse and fine aggregates typically have a dielectric permittivity in the range of $\varepsilon_S = 4\text{-}7$ (Tsui and Matthews, 1997).

Applying Eq. (18), the function of sensitivity $S_\varepsilon^{\theta_W}$ for volumetric water content $\theta_W$ relative to permittivity $\varepsilon$,

$$S_\varepsilon^{\theta_W}(t,s,\omega,\varepsilon) = \frac{\partial \theta_W(t,s,\omega,\varepsilon)}{\partial \varepsilon} \cdot \frac{\varepsilon}{\theta_W(t,s,\omega,\varepsilon)}, \tag{19}$$

assumes the following expression

$$S_\varepsilon^{\theta_W}(t,s,\omega,\varepsilon) = 2\varepsilon\left[\frac{1}{2\varepsilon+\varepsilon_W(t,s,\omega)} - \frac{1}{2\varepsilon+\varepsilon_S}\right] + \frac{2\varepsilon^2+\varepsilon_A\varepsilon_S}{(2\varepsilon+\varepsilon_A)(\varepsilon-\varepsilon_S)-3\varepsilon(\varepsilon_A-\varepsilon_S)\eta}, \tag{20}$$

which can be simplified as:

$$S_\varepsilon^{\theta_W}(t,s,\omega,\varepsilon) \stackrel{\varepsilon_A \ll \varepsilon \ll \varepsilon_W}{\cong} -1 + 2\varepsilon[\frac{1}{\varepsilon_W(t,s,\omega)} + \frac{1}{2\varepsilon - 2\varepsilon_S - 3(\varepsilon_A - \varepsilon_S)\eta}]. \quad (21)$$

The sensitivity function $S_\varepsilon^{\theta_W}$ [Eqs. (20)-(21)] of water content $\theta_W$ relative to $\varepsilon$ shows two asymptotes, the first one horizontal,

$$\lim_{\varepsilon \to \infty} S_\varepsilon^{\theta_W}(t,s,\omega,\varepsilon) = 1, \quad (22)$$

the second one vertical,

$$\lim_{\varepsilon \to \varepsilon^{(asymp)}} S_\varepsilon^{\theta_W}(t,s,\omega,\varepsilon) = +\infty \quad \Rightarrow \quad \varepsilon^{(asymp)} = \varepsilon_S - \frac{3}{2}(\varepsilon_S - \varepsilon_A) \cdot \eta, \quad (23)$$

and one 'knee point' that coincides with the absolute minimum,

$$\left. \frac{\partial S_\varepsilon^{\theta_W}}{\partial \varepsilon}(t,s,\omega,\varepsilon) \right|_{\varepsilon^{(knee)}} = 0 \quad \Rightarrow \quad \varepsilon^{(knee)} \cong \frac{1}{2}\sqrt{\varepsilon_W[3\varepsilon_A \eta + \varepsilon_S(2 - 3\eta)]}. \quad (24)$$

Some points appear worth noting:

- The mathematical–physical model [Eq. (18)] describing the dielectric properties of concrete shows that the volumetric water content $\theta_W(t,s,\omega,\varepsilon)$, which is a function of the relative dielectric permittivity $\varepsilon$, shows almost no dependence on frequency $\omega$, salinity $s$, and temperature $t$, because the dielectric permittivity values for concrete $\varepsilon$, $\varepsilon_S$ are much lower than the permittivity of water, $\varepsilon$, $\varepsilon_S \ll \varepsilon_W(t,s,\omega)$ (Fig. 2a).

- The functions of sensitivity $S_\varepsilon^{\theta_W}$ [Eqs. (20)-(21)] for the volumetric water content $\theta_W$ relative to permittivity $\varepsilon$ depends on frequency $\omega$ to a minor extent, but has almost no dependence on both temperature $t$ and salinity $s$, especially under operating conditions such that $\sigma_W/\omega\varepsilon_0 << \varepsilon_W$.

- The sensitivity function $S_\varepsilon^{\theta_W}$ [Eqs. (20)-(21)] diverges to infinity and is minimized into the values $\varepsilon^{(asym)}$ [Eq. (23)] and $\varepsilon^{(knee)}$ [Eq. (24)], which are functions of the porosity $\eta$ for concrete, and of both the dielectric permittivity values $\varepsilon_A$ and $\varepsilon_S$ for air and the solid components; furthermore, the value $\varepsilon^{(knee)}$ also depends on water permittivity $\varepsilon_W$.

*Terrestrial soil.* For the hypothesis that $\sigma_W/\omega\varepsilon_0 << \varepsilon_W$, Robinson et al. (2002) settled for just an implicit transcendental equation that involves the volumetric content $\theta_W$ of water and the relative dielectric permittivity $\varepsilon$, which solves for a system of equations similar to [Eqs. (1) and (8)]:

$$\begin{cases} \sqrt{\varepsilon} = (1-\eta)\sqrt{\varepsilon_S} + \theta_W\sqrt{\langle\varepsilon_W\rangle} + (\eta-\theta_W)\sqrt{\varepsilon_A} \\ \langle\varepsilon_W\rangle = \dfrac{\varepsilon_W^{(up)}\theta_W/\theta_{BW}}{\ln[1+\dfrac{\varepsilon_W^{(up)}}{\varepsilon_W^{(low)}}(e^{\theta_W/\theta_{BW}}-1)]} \end{cases}. \qquad (25)$$

In addition, below the limit of:

$$\frac{\theta_W}{\theta_{BW}} << 6\frac{\varepsilon_W^{(low)}}{\varepsilon_W^{(low)}+\varepsilon_W^{(up)}}, \qquad (26)$$

the present report proposes an explicit algebraic solution of the equation system [Eq. (25)]; i.e., $\theta_W$ as function of $\varepsilon$ (Fig. 2b):

$$\theta_W(t,s,\varepsilon) \cong 2\theta_{BW} \frac{\sqrt{\varepsilon_W^{(low)}}[\sqrt{\varepsilon_A}-\sqrt{\varepsilon_W^{(low)}}]}{\varepsilon_W^{(up)}(t,s)-\varepsilon_W^{(low)}} \cdot$$
$$\cdot \left\{ 1 - \sqrt{1 + \frac{1}{\theta_{BW}} \frac{[\varepsilon_W^{(up)}(t,s)-\varepsilon_W^{(low)}][\sqrt{\varepsilon}-\eta\sqrt{\varepsilon_A}-(1-\eta)\sqrt{\varepsilon_S}]}{\sqrt{\varepsilon_W^{(low)}}[\sqrt{\varepsilon_A}-\sqrt{\varepsilon_W^{(low)}}]^2}} \right\}, \quad (27)$$

taking $\varepsilon_S = 3.9$ for the solid phase of mineral soils, and $\varepsilon_S = 5.0$ for the solid phase of organic soils (Roth et al., 1990).

Applying Eq. (27), the function of sensitivity $S_\varepsilon^{\theta_W}$ for volumetric water content $\theta_W$ relative to dielectric permittivity $\varepsilon$ assumes the following expression:

$$S_\varepsilon^{\theta_W}(t,s,\varepsilon) = \frac{\partial \theta_W(t,s,\varepsilon)}{\partial \varepsilon} \cdot \frac{\varepsilon}{\theta_W(t,s,\varepsilon)} \cong$$
$$\cong \frac{1}{4} \frac{\frac{K_W^{(up)}(t,s)}{K_A^2}\sqrt{\varepsilon}}{\sqrt{1+\frac{K_W^{(up)}(t,s)}{K_A^2}[\sqrt{\varepsilon}-\eta\sqrt{\varepsilon_A}-(1-\eta)\sqrt{\varepsilon_S}]}\left\{1+\frac{K_W^{(up)}(t,s)}{K_A^2}[\sqrt{\varepsilon}-\eta\sqrt{\varepsilon_A}-(1-\eta)\sqrt{\varepsilon_S}]\right\}}$$

,

(28)

when:

$$K_W^{(up)}(t,s) = \theta_{BW}\sqrt{\varepsilon_W^{(low)}}[\varepsilon_W^{(up)}(t,s)-\varepsilon_W^{(low)}], \quad (29)$$

$$K_A = \theta_{BW} \sqrt{\varepsilon_W^{(low)}} [\sqrt{\varepsilon_A} - \sqrt{\varepsilon_W^{(low)}}]. \tag{30}$$

The sensitivity function $S_\varepsilon^{\theta_W}$ [Eq. (28)] of water content $\theta_W$ relative to permittivity $\varepsilon$ shows one horizontal asymptote,

$$\lim_{\varepsilon \to \infty} S_\varepsilon^{\theta_W}(t,s,\omega,\varepsilon) = 1/4, \tag{31}$$

and one absolute maximum,

$$S_\varepsilon^{\theta_W}\Big|^{(max)} = \frac{1}{4} \frac{K_W^{(up)}(t,s)}{K_A^2} \sqrt{\varepsilon^{(max)}}, \tag{32}$$

$$\sqrt{\varepsilon^{(max)}} = \eta \sqrt{\varepsilon_A} + (1-\eta)\sqrt{\varepsilon_S}, \tag{33}$$

such that, in the trivial case:

$$\lim_{\eta \to 1} \varepsilon^{(max)} = \varepsilon_A. \tag{34}$$

Some points are worth noting:

- When the water phase [see section 3] analyzed in the HF band ($\omega_0 = 2\pi f_0$, $f_0 < 1$ GHz), is characterized by low salinity $s_{low}$ ($s_{low} \to 1$ ppt), temperature $t$ has almost no influence on the measurements of the relative dielectric permittivity values $\varepsilon_W(t,s,\omega)$ for water, and $\varepsilon(t,s,\omega)$ for terrestrial soils, and so on their volumetric water content $\theta_W(\varepsilon)$ [Eq. (27)].

For each non-saturated soil variety ($\alpha \to 0$) (section 3), the frequency $\omega$ influences the salinity $s(\sigma)$ measurements, but not the volumetric water content $\theta_W(\varepsilon)$ measurements [Eq. (27)], because even for HFs, water is characterized by an electrical conductivity $\sigma(t,s,\omega)$ which varies quadratically with $\omega$,

$$\sigma_W(t,s,\omega) \cong \sigma_{stat}(t,s) + \omega^2 \varepsilon_0 \tau [\varepsilon_{stat}(t,s) - \varepsilon_\infty]$$

and a dielectric permittivity $\varepsilon_W(t,s,\omega)$ that remains constant with $\omega$,

$$\varepsilon_W(t,s,\omega) \cong \varepsilon_{stat}(t,s).$$

- The function of sensitivity $S_\varepsilon^{\theta_W}$ [Eq. (28)] for water content $\theta_W$ relative to permittivity $\varepsilon$ has minimal dependence on frequency $\omega$, especially under operating conditions such that $\sigma_W/\omega\varepsilon_0 \ll \varepsilon_W$.
- The sensitivity function $S_\varepsilon^{\theta_W}$ [Eq. (28)] is maximized into the refractive index value $\sqrt{\varepsilon^{(max)}}$ [Eq. (33)], which is a linear combination of the refractive indices $\sqrt{\varepsilon_A}$ and $\sqrt{\varepsilon_S}$ for air and the solid components of soil, respectively, and it depends on their porosity $\eta$.